\documentclass[twocolumn,showpacs,prl]{revtex4}
\usepackage{mathrsfs}
\usepackage{longtable,lscape}
\usepackage{multirow}
\usepackage{txfonts}
\usepackage{amssymb}
\usepackage{indentfirst}
\usepackage{graphicx,,booktabs}
\usepackage{color}
\usepackage{amssymb}

\begin{document}
\title{$Z_b(10610)$ and $Z_b(10650)$ structures produced by the initial single pion emission in the $\Upsilon(5S)$ decays}
\author{Dian-Yong Chen$^{1,3}$}
\author{Xiang Liu$^{1,2}$\footnote{Corresponding author}}\email{xiangliu@lzu.edu.cn}
\affiliation{
$^1$Research Center for Hadron and CSR Physics,
Lanzhou University and Institute of Modern Physics of CAS, Lanzhou 730000, China\\
$^2$School of Physical Science and Technology, Lanzhou University, Lanzhou 730000,  China\\
$^3$Nuclear Theory Group, Institute of Modern Physics of CAS, Lanzhou 730000, China}
\date{\today}

\begin{abstract}
We propose a unique mechanism called {\it Initial Single Pion
Emission} existing in the $\Upsilon(5S)$ decays, and further study
the line shapes of $d\Gamma(\Upsilon(5S\to
\Upsilon(nS)\pi^+\pi^-))/dm_{\Upsilon(nS)\pi^+}$ ($n=1,2,3$) and
$d\Gamma(\Upsilon(5S\to h_b(mP)\pi^+\pi^-))/dm_{h_b(mP)\pi^+}$
($m=1,2$). We find sharp structures around 10610 MeV and 10650 MeV
in the obtained theoretical line shapes of $d\Gamma(\Upsilon(5S\to
\Upsilon(nS)\pi^+\pi^-))/dm_{\Upsilon(nS)\pi^+}$ and
$d\Gamma(\Upsilon(5S\to h_b(mP)\pi^+\pi^-))/dm_{h_b(mP)\pi^+}$
distributions, which could naturally correspond to the
$Z_b(10610)$ and $Z_b(10650)$ structures newly observed by Belle.
\end{abstract}

 \maketitle
%\end{CJK}

\begin{center}
\begin{figure}[htb]
\scalebox{0.64}{\includegraphics{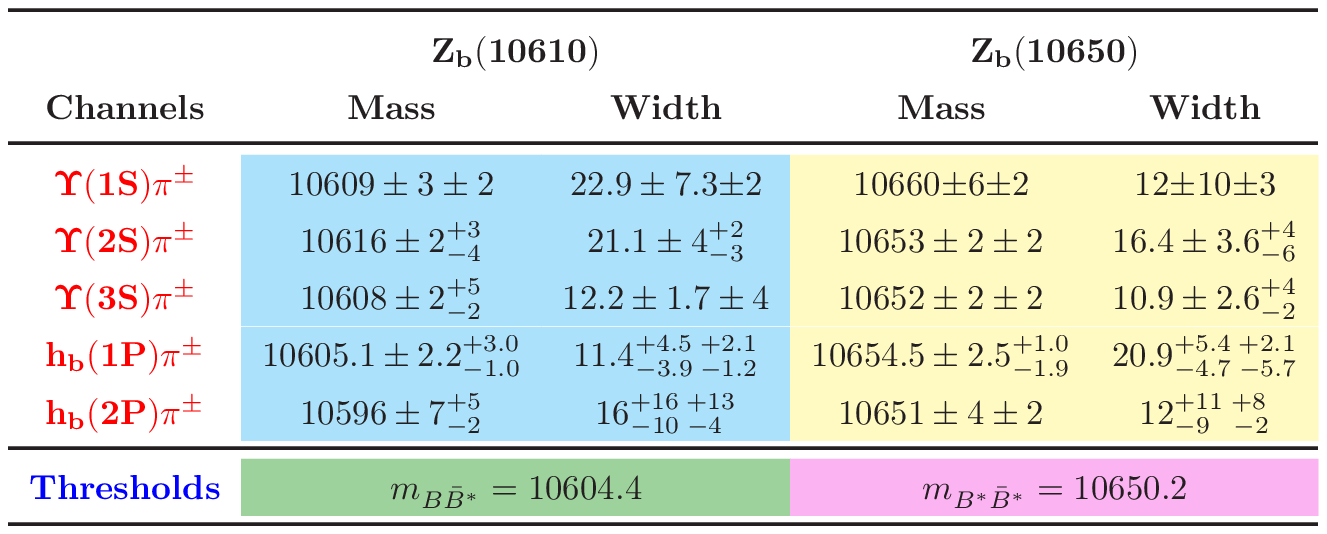}} \caption{(Color online.) The measured
parameters of $Z_b(10610)$ and $Z_b(10650)$ by five different decay
channels \cite{Collaboration:2011gj}, and the comparison of these
parameters with the $B\bar{B}^*$ and $B^*\bar{B}^*$ threshold
\cite{Nakamura:2010zzi}. Here, all values are in units of MeV.
\label{RP}}
\end{figure}
\end{center}

As the first experimental observation of charged bottomonium-like
states, two structures $Z_b(10610)$ and $Z_b(10650)$ were reported
by the Belle Collaboration recently by studying the invariant mass
spectra of $\Upsilon(nS)\pi^\pm$ $(n=1,2,3)$ and $h_b(mP)\pi^\pm$
$(m=1,2)$ of $\Upsilon(5S)\to \Upsilon(nS)\pi^\pm$ and
$\Upsilon(5S)\to h_b(mP)\pi^\pm$ decay processes
\cite{Collaboration:2011gj}. The average values of the mass and the
width of $Z_b(10610)$ and $Z_b(10650)$ are
$M_{Z_b(10610)}=10608.4\pm2.0$ MeV/c$^2$,
$\Gamma_{Z_b(10610)}=15.6\pm2.5$ MeV/c$^2$,
$M_{Z_b(10650)}=10653.2\pm1.5$ MeV/c$^2$,
$\Gamma_{Z_b(10650)}=14.4\pm3.2$ MeV/c$^2$
\cite{Collaboration:2011gj}. In Fig. \ref{RP}, we also list
different measurement results of the parameters of $Z_b(10610)$ and
$Z_b(10650)$ extracted from their five hidden-bottom decay channels,
and compare these results with the $B\bar{B}^*$ and $B^*\bar{B}^*$
thresholds. In addition, Belle also
indicated that $Z_b(10610)$ and $Z_b(10650)$ favor
$I^G(J^{P})=1^+(1^+)$ from the analysis of angular distribution.

The above experimental information shows that $Z_b(10610)$ and
$Z_b(10650)$ are very peculiar since $Z_b(10610)$ and $Z_b(10650)$
not only are charged structures but also close to the thresholds of
$B\bar{B}^*$ and $B^*\bar{B}^*$ respectively. Thus, the $Z_b(10610)$
and $Z_b(10650)$ structures enrich the observation of
bottomonium-like states, and inspire theorists' extensive
interest in revealing what is the source to generate these novel
structures at the same time.

Their peculiarities make that $Z_b(10610)$ and $Z_b(10650)$ could be
as good candidate of exotic states, i.e., $B\bar{B}^*$ and
$B^*\bar{B}^*$ molecular states, which were suggested in Refs.
\cite{Liu:2008fh,Liu:2008tn}. After finding $Z_b(10610)$ and
$Z_b(10650)$ structures, many theoretical work has focused on this
hot issue of $Z_b(10610)$ and $Z_b(10650)$. In Ref.
\cite{Bondar:2011ev}, the decay behavior of $Z_b(10610)$ and
$Z_b(10650)$ was discussed by the heavy quark symmetry and the assignment of
$Z_b(10610)$ and $Z_b(10650)$ as the $J=1$ S-wave $B\bar{B}^*$ and
$B^*\bar{B}^*$ molecular states. Chen, Liu and Zhu
\cite{Chen:2011zv} found that introducing the intermediate
$Z_b(10610)$ and $Z_b(10650)$ contributions to $\Upsilon(5S)\to
\Upsilon(2S)\pi^+\pi^-$ naturally explains Belle's previous
observation of the anomalous $\Upsilon(2S)\pi^+\pi^-$ production
near the peak of $\Upsilon(5S)$ at $\sqrt s=10.87$ GeV
\cite{Abe:2007tk}. By the QCD sum rule and the constructed
$B\bar{B}^*$ molecular current, the authors in Ref.
\cite{Zhang:2011jj} reproduced the mass of $Z_b(10610)$. In Ref.
\cite{Yang:2011rp}, the mass spectra of the S-wave
$[\bar{b}q][b\bar{q}]$, $[\bar{b}q]^*[b\bar{q}]$,
$[\bar{b}q]^*[b\bar{q}]^*$ were calculated in the chiral quark
model, which indicates that $Z_b(10610)$ and $Z_b(10650)$ could be
as the S-wave $B\bar{B}^*$ and $B^*\bar{B}^*$ molecular states. Bugg
proposed that $Z_b(10610)$ and $Z_b(10650)$ are from the cusp effect
due to the $B\bar{B}^*$ and $B^*\bar{B}^*$ thresholds
\cite{Bugg:2011jr}. Two positive C-parity isoscalar states, i.e., a
$^3S_1-^3D_1$ state with a binding energy of 90-100 MeV and a
$^3P_0$ state located about 20-30 MeV below the $B\bar{B}^*$
threshold, were suggested in Ref. \cite{nieves}. However, the
quantum numbers relevant to these suggested molecular bottomonia are
inconsistent with those of the observed two charged $Z_b$ states.
The authors in Ref. \cite{russian} studied the interaction between a
light hadron and heavy quarkonium through the transition to a pair
of intermediate heavy mesons, and discussed the resonance structures
close to the $B^{(*)}\bar B^\ast$ threshold \cite{russian}. In Ref.
\cite{Guo:2011gu}, the discussion of $Z_b(10610)$ and $Z_b(10650)$
being tetraquark states was performed by using the chromomagnetic
interaction. We notice that the $b\bar b q\bar q$ tetraquark states
with $10.2 \sim 10.3$ GeV were once predicted in Ref. \cite{cui} by
using the color-magnetic interaction with the flavor symmetry
breaking corrections, where the predicted mass is lower than that
obtained in Ref. \cite{Guo:2011gu} and consistent with the values
extracted from the QCD sum rule \cite{chenwei}. Very recently, the
interactions of the $B^*\bar{B}$ and $B^*\bar{B}^{*}$ were revisited
by the one-boson-exchange model. After considering the S-wave and
D-wave mixing, we notice that both $Z_b(10610)^\pm$ and
$Z_b(10650)^\pm$ can be interpreted as the $B^*\bar{B}$ and
$B^*\bar{B}^{*}$ molecular states \cite{Sun:2011uh}.

Generally speaking, these theoretical efforts mentioned above have
improved our understanding of the properties of $Z_b(10610)$ and
$Z_b(10650)$, especially stimulated the extensive and
in-depth study of exotic states, which is an important and valuable
research topic in hadron physic at present. If revealing the underlying mechanism behind these novel $Z_b$
structures much more
comprehensively, we need to pay more phenomenological efforts from
different perspectives. Thus, the study of whether $Z_b(10610)$ and
$Z_b(10650)$ can be depicted without introducing any exotic
structure explanation is becoming a very valuable research issue.
Along this way, we will delve into this subject.

With $\Upsilon(5S)\to \Upsilon(nS)\pi^+\pi^-$ as an example, we
first illustrate the corresponding decay mechanisms of the
hidden-bottom decays of $\Upsilon(5S)$. One is that $\Upsilon(5S)$
directly decays into $\Upsilon(nS)\pi^+\pi^-$, which is usually
depicted by the QCD Multipole Expansion method
\cite{Kuang:1981se,Yan:1980uh,Novikov:1980fa}. Another one is that
the dipion in the $\Upsilon(5S)\to \Upsilon(nS)\pi^+\pi^-$ process could
be from the intermediate states $\sigma(600)$, $f_0(980)$ and
$f_2(1270)$ just indicated in Ref. \cite{Chen:2011qx}, where the
intermediate hadronic loops constructed by the $B^{(*)}$ mesons play
an important role to connect the initial $\Upsilon(5S)$ with the final
$\Upsilon(nS)\pi^+\pi^-$.

\begin{center}
\begin{figure}[htb]
\begin{tabular}{cccc}
\scalebox{0.7}{\includegraphics{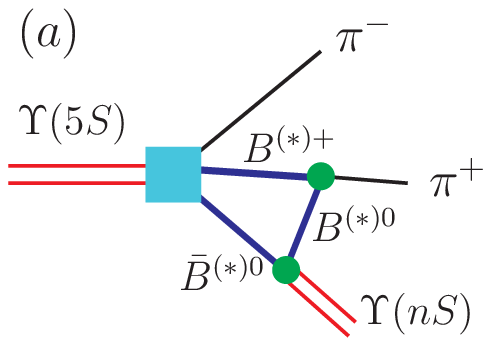}}&\raisebox{3em}{\huge{$+$}}&
\scalebox{0.7}{\includegraphics{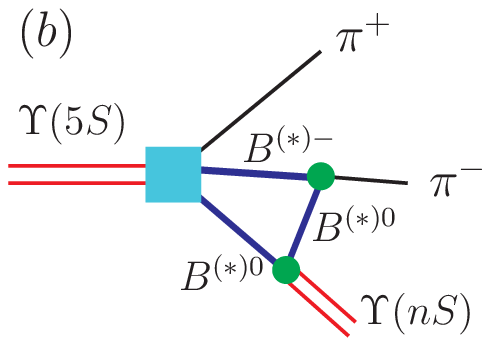}}
\end{tabular}
\caption{(Color online.) The schematic diagrams for $\Upsilon(5S)\to
\Upsilon(nS)\pi^+\pi^-$ by the ISPE mechanism. Here, diagrams (a)
and (b) are related to each other by particle antiparticle conjugation, i.e.,
$B^{(*)}\rightleftharpoons\bar{B}^{(*)}$ and $\pi^+\rightleftharpoons \pi^-$. After performing the transformations $B^{(*)+}\rightleftharpoons B^{(*)0}$, $B^{(*)-}\rightleftharpoons \bar{B}^{(*)0}$ and $\pi^+\rightleftharpoons \pi^-$, we obtain the remaining
diagrams. By replacing $\Upsilon(nS)$ with $h_b(mP)$, one obtains
the diagrams for $\Upsilon(5S)\to h_b(mP)\pi^+\pi^-$. \label{IPE}}
\end{figure}
\end{center}

Besides these two production mechanisms, in this work we propose an
important mechanism contributing to the
$\Upsilon(5S)\to\Upsilon(nS)\pi^+\pi^-$ decay, which is described in
Fig. \ref{IPE}. $\Upsilon(5S)$ transits into $B^{(*)}$ and
$\bar{B}^{(*)}$ pair associated with a single pion emission. Due to
the emitted pion with continuous energy distribution, $B^{(*)}$ and
$\bar{B}^{(*)}$ mesons with the low momentum easily interact with
each other and further transit into $\Upsilon(nS)\pi$ by exchanging
$B^{(*)}$ meson. We name such new picture presented here as
{\it Initial Single Pion Emission} (ISPE) mechanism. 
To some extent,
the ISPE mechanism existing in the $\Upsilon$ decays is similar to
the well-known {\it Initial State Radiation} (ISR) mechanism in $e^+e^-$
collisions, which has stimulated a series of observations of
charmonium-like states $X$, $Y$, $Z$ in the past years.

The ISPE mechanism exists in the hidden-charm or hidden-bottom dipion decays of
higher charmonia or bottomonia. If the mass of higher charmonium/bottomonium is larger than the sum of the masses of $D^{(*)}\bar{D}^{(*)}/B^{(*)}\bar{B}^{(*)}$ pair and pion, this higher charmonium/bottomonium can be of open-charm/open-bottom decays
associated with a pion production. The emitted single pion plays important role to make $D^{(*)}\bar{D}^{(*)}/B^{(*)}\bar{B}^{(*)}$ with low momenta.
Then, $D^{(*)}\bar{D}^{(*)}/B^{(*)}\bar{B}^{(*)}$ into final states occurs via $D^{(*)}/B^{(*)}$ meson exchanges. Thus, under the ISPE mechanism, the hidden-charm/hidden-bottom dipion decays of higher charmonium/bottomonium are mediated by the hadronic loop constructed by $D^{(*)}/B^{(*)}$ 
and $\bar{D}^{(*)}/\bar{B}^{(*)}$ mesons. Since in fact hadronic loop effect reflected the coupled channel effect, the ISPE mechanism can be categorized as an important nonperturbative QCD effect.

Since two $Z_b$ structures were observed in the hidden-bottom decays
of $\Upsilon(5S)$, we naturally relate the newly observed structures
with the $\Upsilon(5S)$ decay via the ISPE mechanism, and further exam
whether the $Z_b$ structures can be reproduced in the
$\Upsilon(nS)\pi^\pm$ invariant mass spectrum when including the
diagrams in Fig. \ref{IPE}.

In the following, we calculate the $\Upsilon(nS)\pi^\pm$ invariant
mass spectra of $\Upsilon(5S)\to
\Upsilon(nS)(p_1)\pi^+(p_2)\pi^-(p_3)$ considering the intermediate
$B\bar{B}$, $B\bar{B}^*+h.c.$ and $B^*\bar{B}^*$ contributions. Just
shown in Fig. \ref{IPE}, the schematic diagrams (a) and (b)
correspond to $\Upsilon(5S)\to
\pi^-+[B^{(*)+}\bar{B}^{(*)0}\rightarrowtail
\Upsilon(nS)\pi^+]_{B^{(*)0}}$ and $\Upsilon(5S)\to
\pi^++[B^{(*)-}{B}^{(*)0}\rightarrowtail
\Upsilon(nS)\pi^-]_{B^{(*)0}}$, where the subscript $B^{(*)0}$ denotes
the exchanged meson for
$B^{(*)}\bar{B}^{(*)}\rightarrowtail\Upsilon(nS)\pi$ transitions.
Thus, there exist two, six and four independent decay amplitudes for the
$\Upsilon(5S)\to \Upsilon(nS)\pi^+\pi^-$ decays via the intermediate
$B\bar{B}$, $B\bar{B}^*+h.c.$ and $B^*\bar{B}^*$ respectively.

The general expressions corresponding to Fig. \ref{IPE} (a) and (b)
can be written as
\begin{eqnarray}
&&\mathcal{M}\big\{\Upsilon(5S)\to \pi^-+[B^{(*)+}\bar{B}^{(*)0}\rightarrowtail\Upsilon(nS)\pi^+]_{B^{(*)0}}\big\}\nonumber\\
&&=\prod_i g_i\int\frac{d^4 q}{(2\pi)^4}\frac{\big[p_1,p_2,p_3,q\big]_{\mu\nu}\epsilon_{\Upsilon(5S)}^{\mu}
\epsilon_{\Upsilon(nS)}^{\nu}}{\big[(p_2+q)^2-m_{B^{(*)}}^2\big]\big[(p_1-q)^2-m_{B^{(*)}}^2\big]}\nonumber\\&&\quad\times
\frac{1}{q^2-m_{B^{(*)}}^2}\mathcal{F}^2(q^2,m_{B^{(*)}}^2),\label{h1}\\
&&\mathcal{M}\big\{\Upsilon(5S)\to \pi^++[B^{(*)-}{B}^{(*)0}\rightarrowtail\Upsilon(nS)\pi^-]_{B^{(*)0}}\big\}\nonumber\\
&&=\prod_i g_i\int\frac{d^4
q}{(2\pi)^4}\frac{\big[p_1,p_2,p_3,q\big]_{\mu\nu}\epsilon_{\Upsilon(5S)}^{\mu}
\epsilon_{\Upsilon(nS)}^{\nu}}{\big[(p_3+q)^2-m_{B^{(*)}}^2\big]\big[(p_1-q)^2-m_{B^{(*)}}^2\big]}\nonumber\\&&\quad\times
\frac{1}{q^2-m_{B^{(*)}}^2}\mathcal{F}^2(q^2,m_{B^{(*)}}^2),\label{h2}
\end{eqnarray}
where $\big[p_1,p_2,p_3,q\big]_{\mu\nu}$ denotes the Lorentz
structures constructed by four-momenta $p_1$, $p_2$, $p_3$ and $q$,
which are obtained by the effective Lagrangian approach \cite{Oh:2000qr,Casalbuoni:1996pg,Colangelo:2003sa}
\begin{eqnarray}
%%
%%Upsilon B(*) B(*) \pi
&&\mathcal{L}_{\Upsilon B^{(*)} B^{(*)} \pi}\nonumber\\&&=-ig_{\Upsilon BB \pi}
\varepsilon^{\mu \nu \alpha \beta} \Upsilon_{\mu} \partial_{\nu} B
\partial_{\alpha} \pi \partial_{\beta} \bar{B} + g_{\Upsilon B^\ast B \pi} \Upsilon^{\mu} (B \pi
\bar{B}^\ast_{\mu} + B^\ast_{\mu} \pi \bar{B}) \nonumber\\
&&\quad-ig_{\Upsilon B^\ast B^\ast \pi} \varepsilon^{\mu \nu \alpha
\beta} \Upsilon_{\mu} B^\ast_{\nu} \partial_{\alpha} \pi
\bar{B}^\ast_\beta -ih_{\Upsilon B^\ast B^\ast \pi} \varepsilon^{\mu \nu \alpha
\beta} \partial_{\mu} \Upsilon_{\nu} B^\ast_{\alpha} \pi
\bar{B}^\ast_{\beta},\nonumber\\
%%
%%B*B(*) pi
&&\mathcal{L}_{B^\ast B^{(\ast)} \pi} \nonumber\\&&= ig_{B^\ast B \pi}
(B^\ast_{\mu} \partial^\mu \pi \bar{B}-B \partial^\mu \pi
\bar{B}^\ast_{\mu})-g_{B^\ast B^\ast \pi} \varepsilon^{\mu \nu \alpha \beta}
\partial_{\mu} B^\ast_{\nu} \pi \partial_{\alpha}
\bar{B}^\ast_{\beta},\nonumber\\
%%
%%Upsilon B(*)B(*)
&&\mathcal{L}_{\Upsilon(nS) B^{(*)} B^{(*)}}\nonumber\\&&=ig_{\Upsilon BB}
\Upsilon_{\mu} (\partial^\mu B \bar{B}- B \partial^\mu
\bar{B})-g_{\Upsilon B^\ast B} \varepsilon^{\mu \nu \alpha \beta}
\partial_{\mu} \Upsilon_{\nu} (\partial_{\alpha} B^\ast_{\beta} \bar{B}
\nonumber\\&&\quad+ B \partial_{\alpha} \bar{B}^\ast_{\beta})-ig_{\Upsilon B^\ast B^\ast} \big\{ \Upsilon^\mu (\partial_{\mu}
B^{\ast \nu} \bar{B}^\ast_{\nu} -B^{\ast \nu} \partial_{\mu}
\bar{B}^\ast_{\nu}) \nonumber\\
&&\quad+ (\partial_{\mu} \Upsilon_{\nu} B^{\ast \nu} -\Upsilon_{\nu}
\partial_{\mu} B^{\ast \nu}) \bar{B}^{\ast \mu} + B^{\ast \mu}(\Upsilon^\nu \partial_{\mu} \bar{B}^\ast_{\nu} -
\partial_{\mu} \Upsilon^\nu \bar{B}^\ast_{\nu})\big\},\nonumber\\
&&\mathcal{L}_{h_b(mP) B^{(*)} B^{(*)}}\nonumber\\&&= g_{h_b B^\ast B} h_b^\mu
(\bar{B}^\ast_{\mu} B + B^\ast_\mu \bar{B})+ ig_{h_b B^\ast B^\ast} \varepsilon^{\mu \nu \alpha \beta}
\partial_{\mu} h_{b \nu} B^\ast_{\alpha} \bar{B}^\ast_{\beta}. \nonumber
\end{eqnarray}
In the heavy quark limit, the coupling constants in the above
Lagrangians satisfy the relations $g_{B^\ast B^\ast \pi} =g_{B^\ast
B \pi}/\sqrt{m_B m_B^\ast} =2g/f_{\pi}$, $g_{\Upsilon B B}
=g_{\Upsilon B^\ast B^\ast}=m_{\Upsilon} g_{\Upsilon B^\ast B^\ast}
=m_{\Upsilon}/f_{\Upsilon}$, $g_{h_b B B^\ast} =-2g_1 \sqrt{m_{h_b}
m_B m_{B^\ast}}, \ g_{h_b B^\ast B^\ast} =2g_1 m_B^\ast
/\sqrt{m_{h_b}}$ with $g=0.59$ \cite{Isola:2003fh}, $f_\pi=132$
MeV and $g_1=-\sqrt{m_{\chi_{b0}}/3}/f_{\chi_{b0}}$
\cite{Colangelo:2003sa}, where $f_{\Upsilon}$ and $f_{\chi_{b0}}$
denote the decay constants of $\Upsilon(nS)$ and $\chi_{b0}$. The
mass parameters of $B^{(*)}$, $\Upsilon(5S)$, $\Upsilon(nS)$,
$h_b(mP)$ are taken from Refs.
\cite{Nakamura:2010zzi,Adachi:2011ji}.

In Eqs. (\ref{h1})-(\ref{h2}), $\prod_i g_i$ denotes the product of all coupling constants involving
in three interaction vertices (see Fig. \ref{IPE}). Additionally, we
introduce monopole form factor
$\mathcal{F}(q^2,m_{B^{(*)}}^2)=(\Lambda^2-m_{B^{(*)}}^2)/(\Lambda^2-q^2)$
reflecting the structure effect of the interaction vertices
of $B^{(*)}\bar{B}^{(*)}\rightarrowtail\Upsilon(nS)\pi$ transitions
in $\Upsilon(5S) \to \Upsilon(nS) \pi^+ \pi^-$ decay, which also
compensates the off-shell effects of the mesons at the vertices.
Here, $q$ denotes the four-momentum of the exchanged $B^{(*)}$ meson.
$\Lambda$ is a phenomenological parameter, which can be
parameterized as $\Lambda=m_{B^{(*)}}+\beta \Lambda_{QCD}$ with
$\Lambda_{QCD}=220$ MeV.

The differential decay width for $\Upsilon(5S)
\to \Upsilon(nS) \pi^+ \pi^-$ is,
\begin{eqnarray}
d\Gamma =\frac{1}{3} \frac{1}{(2 \pi)^3} \frac{1}{32
m_{\Upsilon(5S)}^3} \overline{|\mathcal{M}_{\mathrm{total}}^2|}
dm_{\Upsilon(nS) \pi^+}^2 dm_{\pi^+\pi^-}^2
\end{eqnarray}
with $m_{\Upsilon(nS) \pi^+}^2 = (p_1 + p_2)^2$ and
$m_{\pi^+\pi^-}^2 =(p_2 +p_3)^2$, where the overline indicates the
average over the polarizations of the $\Upsilon(5S)$ in the
initial state and the sum over the polarization of $\Upsilon(2S)$ in
the final state.

\begin{center}
\begin{figure}[htb]
\scalebox{0.56}{\includegraphics{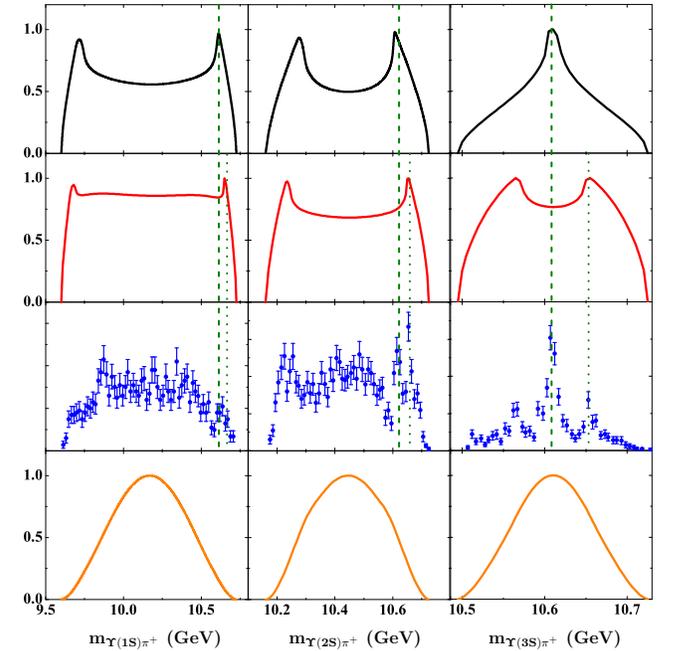}} \caption{(Color
online.) The obtained theoretical line shapes of
$d\Gamma(\Upsilon(5S)\to\Upsilon(nS)\pi^+\pi^-)/d
m_{\Upsilon(nS)\pi^+}$, and the comparison of our result with the
Belle data (the third column) \cite{Collaboration:2011gj}. The first,
the second and the fourth columns correspond to the numerical result
considering $B\bar{B}^*+h.c.$, $B^*\bar{B}^*$ and
$B\bar{B}$ intermediate state contributions respectively, while the
first, the second and the third rows are the results corresponding to
the distributions of the $\Upsilon(1S)\pi^+$, $\Upsilon(2S)\pi^+$
and $\Upsilon(3S)\pi^+$ invariant mass spectra. We use the vertical
dashed and dotted lines to mark the masses of $Z_b(10610)$ and
$Z_b(10650)$, respectively. Here, the maximum of the theoretical
line shape is normalized to 1. \label{UP}}
\end{figure}
\end{center}

With the above preparation, we obtain the line shapes of
$d\Gamma(\Upsilon(5S)\to\Upsilon(nS)\pi^+\pi^-)/d
m_{\Upsilon(nS)\pi^+}$ dependent on the $m_{\Upsilon(nS)\pi^+}$,
which are presented in Fig. \ref{UP}. To explicitly illustrate the
phenomena of the ISPE effect on
$\Upsilon(5S)\to\Upsilon(nS)\pi^+\pi^-$ decays, we individually
consider the intermediate $B\bar{B}^*+h.c.$, $B^*\bar{B}^*$ and
$B\bar{B}$ contributions to
$\Upsilon(5S)\to\Upsilon(nS)\pi^+\pi^-$ process. Thus, we take the
coupling constants of $\Upsilon(5S)$ interacting with
$B^{(*)}\bar{B}^{(*)}\pi$ as 1, which does not change the line
shapes of $d\Gamma(\Upsilon(5S)\to\Upsilon(nS)\pi^+\pi^-)/d
m_{\Upsilon(nS)\pi^+}$. In our calculation, $\beta=1$ is taken. We
need to specify that these line shapes are weakly dependent on the
$\beta$ value, which makes the qualitative conclusion obtained in
this work to be unchanged.

\begin{center}
\begin{figure}[htb]
\scalebox{0.55}{\includegraphics{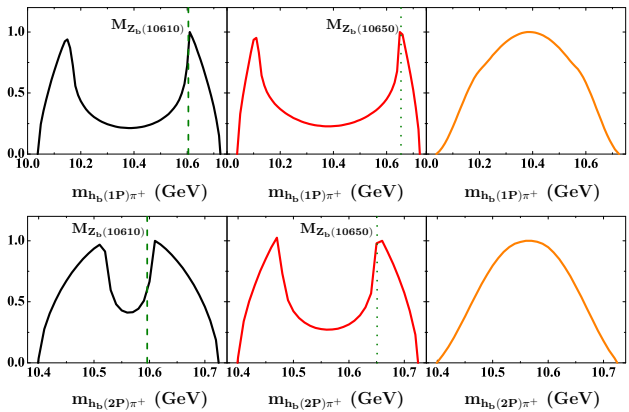}} \caption{(Color online.)
The theoretical curves of $d\Gamma(\Upsilon(5S)\to
h_b(1P)\pi^+\pi^-)/d m_{h_b(1P)\pi^+}$ (the first column) and
$d\Gamma(\Upsilon(5S)\to h_b(2P)\pi^+\pi^-)/d m_{h_b(2P)\pi^+}$ (the
second column). For easily comparing our result with the experimental
data, one adopts the vertical dashed and dotted lines to denote the
masses of $Z_b(10610)$ and $Z_b(10650)$ respectively. The first, the
second and the third rows correspond to the numerical result respectively
considering $B\bar{B}^*+h.c.$, $B^*\bar{B}^*$ and $B\bar{B}$
intermediate state contributions in Fig. \ref{IPE}. Here, the
maximum of the theoretical line shape is normalized to 1.
\label{HB}}
\end{figure}
\end{center}

Just shown in the first and the second columns of Fig. \ref{UP}, combined with the
corresponding reflections, the sharp peaks around $B\bar{B}^*$ and
$B^*\bar{B}^*$ thresholds appear in the $m_{\Upsilon(1S)\pi^+}$ and
$m_{\Upsilon(2S)\pi^+}$ distributions of
$d\Gamma(\Upsilon(5S)\to\Upsilon(1S)\pi^+\pi^-)/d
m_{\Upsilon(1S)\pi^+}$ and
$d\Gamma(\Upsilon(5S)\to\Upsilon(2S)\pi^+\pi^-)/d
m_{\Upsilon(2S)\pi^+}$ \footnote{To some extent, these theoretical curves with sharp peaks presented in Figs. \ref{UP} and \ref{HB}
are shaped similar to a {\it kitty} head.}. The comparison of these results with the
Belle data \cite{Collaboration:2011gj} indicates that we indeed can
mimic the peak structures similar to the $Z_{b}(10610)$ and
$Z_{b}(10650)$ reported by Belle if introducing the ISPE
mechanism.

The theoretical result of the
$\Upsilon(5S)\to\Upsilon(3S)\pi^+\pi^-$ decay further indicates that
there also exists a peak around 10610 MeV, which combines with its
reflection in the $m_{\Upsilon(3S)\pi^+}$ distribution to form a
broad structure. In addition, a structure at $\sim 10650$ MeV and
its reflection are reproduced. These results qualitatively and
naturally explain why there are three structures appearing in the
$\Upsilon(3S)\pi^+$ invariant mass spectrum just announced by Belle
\cite{Collaboration:2011gj}.

We continue to extent the ISPE mechanism to study the
$\Upsilon(5S)\to h_b(mP)\pi^+\pi^-$ ($m=1,2$) decays. Similar to the
situation of $\Upsilon(5S)\to \Upsilon(1S,2S)\pi^+\pi^-$, we can
also find two structures around 10610 MeV and 10650 MeV and their
reflections in the theoretical line shape of $d\Upsilon(5S)\to
h_b(mP)\pi^+\pi^-/dm_{h_b(mP)\pi^+}$, which are consistent with the
Belle's observation well \cite{Collaboration:2011gj}.

The Belle data also give a very intriguing phenomenon, i.e., there
does not exist the structure near the $B\bar{B}$
threshold. Our mechanism can provides a direct explanation to it.
If only considering the $B\bar{B}$ contribution in Fig. \ref{IPE}, our
calculation shows that we cannot find the sharp peak close to the
$B\bar{B}$ threshold in the $\Upsilon(nS)\pi^+$ and $h_b(mP)\pi^+$ invariant mass spectra.
Alternately, the smooth line shapes similar to phase space of
corresponding decay processes appear in the invariant mass spectra
of $\Upsilon(nS)\pi^+$ and $h_b(mP)\pi^+$.

In summary, stimulated by the newly observed two charged $Z_b$
structures \cite{Collaboration:2011gj}, we proposed a new decay
mechanism of $\Upsilon(5S)$, the {\it Initial Single Pion
Emission} mechanism, to study the distributions of the
$\Upsilon(nS)\pi^+$ and $h_b(mP)$ invariant mass spectra in the
$\Upsilon(5S)$ decays into $\Upsilon(nS)\pi^+\pi^-$ and
$h_b(mP)\pi^+\pi^-$. By emitting a pion, $\Upsilon(5S)$ decays
into $B^{(*)}$ and $\bar{B}^{(*)}$ mesons with low momentum, which
can easily interact with each other to transit into
$\Upsilon(nS)\pi^+\pi^-$ or $h_b(mP)\pi^+\pi^-$. The further
calculation shows that there exist sharp structures around 10610
MeV and 10650 MeV in the obtained theoretical line shapes of
$d\Gamma(\Upsilon(5S\to
\Upsilon(nS)\pi^+\pi^-))/dm_{\Upsilon(nS)\pi^+}$ and
$d\Gamma(\Upsilon(5S\to h_b(mP)\pi^+\pi^-))/dm_{h_b(mP)\pi^+}$
distributions. We naturally explain why the Belle Collaboration
can find the charged $Z_b(10610)$ and $Z_b(10650)$ structure in
five different hidden-bottom decay channels. Thus, the ISPE
mechanism presented in this letter provides a unique perspective
to understand the Belle's observation \cite{Collaboration:2011gj}
without introducing any exotic state assignments. Additionally,
our model also answers why Belle did not find the charged
structure near the $B\bar{B}$ threshold in the
$\Upsilon(nS)\pi^+\pi^-$ and $h_b(mP)\pi^+\pi^-$ channels.

If the ISPE mechanism is a universal mechanism existing in the
$\Upsilon(5S)$ decays, this study presented in this letter can be
extended to include the theoretical study of the dipion
hidden-bottom decays of $\Upsilon(11020)$, and even the dipion
hidden-charm decays of higher charmonia $\psi(4040)$, $\psi(4160)$
and $\psi(4415)$, which could produce some other similar
structures near the thresholds of $B_{(s)}^{(*)}$ or
$D_{(s)}^{(*)}$ meson pair. Further experimental search for these
novel phenomenons will be an interesting research topic.

\vfil \noindent {\it Acknowledgment}: This project is supported by
the National Natural Science Foundation of China under Grants Nos.
1175073, No. 11005129, No. 11035006, No. 11047606, the Ministry of
Education of China (FANEDD under Grant No. 200924, DPFIHE under
Grant No. 20090211120029, NCET under Grant No. NCET-10-0442, the
Fundamental Research Funds for the Central Universities), and the
West Doctoral Project of Chinese Academy of Sciences.

\end{document}